\documentclass[pra,twocolumn,amsmath,amssymb,10pt,aps,longbibliography,superscriptaddress,citeautoscript,bibnotes,footinbib]{revtex4-1}
\usepackage{draftcopy}
\usepackage{graphicx}
\usepackage{verbatim}
\usepackage{amsmath}
\usepackage{algpseudocode}
\usepackage{mathrsfs}
\usepackage{lineno}
\usepackage{amssymb}

\usepackage{amsmath}
\usepackage{bm}
\usepackage{latexsym}
\usepackage{graphics}
\usepackage{dsfont}
\usepackage{amssymb}
\usepackage{tabularx}
\usepackage{bbm}
\usepackage{bbold}
\usepackage{tikz}
\usetikzlibrary{quantikz}
\usepackage[colorlinks=true]{hyperref}   
\usepackage[dvipsnames]{xcolor}

\begin{document}

\title{Mitigating Trotter Errors via Post-Processed Symmetry Restoration}

\author{Sangjin Lee}
\thanks{Electronic Address: sangjin5190@gmail.com}
\affiliation{Quantum Universe Center, Korea Institute for Advanced Study, Seoul 02455, South Korea}

\author{Sangkook Choi}
\thanks{Electronic Address: sangkookchoi@kias.re.kr}
\affiliation{School of Computational Sciences, Korea Institute for Advanced Study}
\affiliation{Quantum Universe Center, Korea Institute for Advanced Study, Seoul 02455, South Korea}

\date{\today}
 
\begin{abstract} 
 Quantum simulation is a powerful tool for exploring complex quantum many-body systems such as condensed matter physics and gauge theories. Trotterization, which approximates the ideal time evolution operator by decomposing it into a sequence of local gate operations, is one of the most widely used quantum simulation algorithms. However, such Trotterized implementations generally fail to preserve the symmetries of the target Hamiltonian during compilation. As a result, they can drive quantum states out of {symmetrically allowed} subspaces, leading to unphysical dynamics and symmetry-violating algorithmic errors. In this work, we propose a symmetry-based Trotter error mitigation protocol using classical post-processing. By applying symmetry transformations to the initial state {or interleaving them between discrete Trotter layers}, and then {averaging an ensemble of} the resulting measurement outcomes {via classical post-processing}, our method systematically projects out the symmetry-violating components of the Trotter error while leaving the ideal dynamics unchanged. Importantly, this framework naturally accommodates non-local spatial symmetries and anti-unitary operations such as time reversal, which are difficult or impossible to implement directly with hardware-native {quantum} gates. We benchmark our protocol on the one-dimensional $XY$ model and the one-dimensional Schwinger model. In the $XY$ model, enforcing reflection symmetry suppresses the leading-order Trotter error, whereas in the Schwinger model, {interleaving gauge transformations between Trotter layers enables} gauge-twirling effectively {to reduce} unphysical violations of local Gauss’s law. These results demonstrate that symmetry-based post-processing provides a depth-preserving route to substantially improving the fidelity of Trotterized quantum simulations on near-term devices.
\end{abstract}

\maketitle
\section{introduction}
Symmetry is one of the most fundamental properties of a Hamiltonian, dictating conserved quantities, selection rules, and the structure of quantum dynamics. In many-body systems and quantum field theories, symmetries rigorously constrain the accessible Hilbert space and determine physically allowed processes~\cite{wiger_symmetry,sym2,sym3,sym4,sym5}. These invariant properties therefore provide a natural diagnostic for distinguishing physically admissible dynamics from unphysical symmetry-violating behavior in theoretical analyses and quantum simulations.

To explore the complex dynamics of quantum many-body systems, quantum simulation has emerged as a highly promising tool~\cite{qsim_general,qsim_general2,qsim_general3,LGT_simul,LGT2,qsim_chemistry1,qsim_chemistry2,qsim_cond1,qsim_cond2,qsim_cond3,qsim_cond4}. However, implementing the ideal unitary time evolution operator for a many-body Hamiltonian is generally intractable on current quantum hardware. A widely used approach to this problem is Trotterization, where the time-evolution operator is decomposed into a sequence of hardware-native gates~\cite{Suzuki1,Suzuki2,qsim_appl2,qsim_appl,PF_second,PF_ruth2,PF_ruth,PF_random2,PF_random,PF_chin}. 

In practice, the realization of such Trotterized circuits is severely limited because of restricted qubit connectivity and imperfect gate fidelities, so that suppressing algorithmic errors typically requires prohibitively deep quantum circuits. To address these realistic limitations, a variety of classical post-processing methods have been developed to mitigate algorithmic errors~\cite{QEM_endo,TEP,MPF0,MPF2,MPF3,MPF}, alongside complementary methods aimed at suppressing physical hardware errors~\cite{PEC1,PEC2,QEM_review,ZNE1,ZNE2}. By applying post-processing to the outputs of shallow quantum circuits, one can substantially enhance the performance of quantum simulations within a fixed quantum resource budget, making such strategies particularly well-suited to near-term quantum devices.

Despite the practical utility of Trotterization, a central challenge is that Trotterized circuit does not, in general, respect the symmetries of the target Hamiltonian. 
This discrepancy causes the simulated quantum state to gradually leak into unphysical subspaces of the Hilbert space as the simulation evolves. 
Therefore, preserving these symmetries is critically important across various domains. For example, local gauge invariance enforces strict constraints such as Gauss's law in the quantum simulation of lattice gauge theories~\cite{LGT_review,LGT_simul}, while crystalline symmetries are indispensable in condensed matter systems for protecting gapless Dirac fermions in graphene-like materials and stabilizing robust boundary modes in symmetry-protected topological phases~\cite{SPT,SL}. Symmetry-violating Trotter errors in these simulations drive the system into unphysical states, invalidating the physical interpretation of the computed dynamics~\cite{qsim_algo_sym,gauge_fixing_simul,gauge_oracle,gauge_penalty,gauge_penalty2,gauge_protection,gauge_shearing,gauge_sym_verification}.	

{There have been numerous proposals to protect symmetries in quantum simulations. Representative examples include post-selection schemes, applied either during the Trotter circuit or at its end, which detect and discard states that violate symmetry constraints~\cite{gauge_sym_verification,sym_verification,sym_veri2}. Another well-known strategy is to insert symmetry transformations between Trotter layers, exploiting the fact that such transformations act nontrivially on Trotter errors~\cite{qsim_algo_sym}. While these approaches provide fundamental pathways to symmetry preservation, they naturally entail distinct hardware demands. For instance, post-selection schemes generally necessitate ancilla-assisted measurements or specific readout setups to evaluate the symmetry operators, whereas repeatedly inserting symmetry operations can significantly extend the coherent circuit depth. Thefore, there remains a strong need for hardware-efficient strategies that can robustly preserve symmetries without exacerbating the stringent constraints of near-term quantum devices.}

In this work, we propose a symmetry-preserving Trotter error mitigation method that effectively restores the symmetries violated by Trotterized circuits. Since ideal expectation values are invariant under symmetry transformations, we apply symmetry operations either to the initial state or {interleave them between discrete Trotter layers}. By averaging {an ensemble of} the resulting measurement outcomes through classical post-processing, symmetry-violating contributions to the Trotter error are systematically projected out while the symmetry-preserving part of
the dynamics is retained.

A key advantage of our method is its ability to accommodate non-local symmetries and anti-unitary operations such as time reversal. These transformations are either highly resource-intensive or fundamentally impossible to implement with standard unitary quantum gates. By shifting these demanding symmetry operations to the state preparation stage or to discrete intermediate layers, we effectively bypass these hardware limitations. As a result, our protocol robustly suppresses unphysical leakage and enhances simulation fidelity within realistic circuit-depth and hardware constraints.

The remainder of this paper is organized as follows. In Section~\ref{background}, we review the principles of Trotter-simulation and the impact of symmetry-violating algorithmic errors. In Section~\ref{protocol}, we provide the theoretical framework of symmetry based mitigation using post-processing{, detailing both the initial-state symmetrization and the interleaved-circuit approach}. In Section~\ref{numerics}, we { present} numerical demonstrations of our strategy applied to characteristic quantum many-body systems and gauge theories{, specifically the one-dimensional $XY$ model and the Schwinger model}. Finally, we conclude with a summary and a discussion of future outlooks in Section~\ref{discussion}. In the Appendix, we provide {a} basic review of the Schwinger model (Appendix.~\ref{schwinger_review}), which we used in the numerics section, and a discussion of physical error channel simplification by the symmetrization (Appendix.~\ref{error_simplification}).

\section{Background : Trotter-simulation and Symmetry}
\label{background}
\subsection{Trotter-simulation}
In quantum simulation, a primary task is to obtain the time evolution of a physical observable $\mathcal{O}$ with respect to an initial state $|\psi\rangle$, whose dynamics are governed by a target Hamiltonian $H=\sum_{i} H_i$. Since the ideal time-evolution operator is given by $U(t) = e^{-i H t}$, the ideal expectation value of the observable at time $t$ is written as
\begin{align*}
\langle \mathcal{O}(t)\rangle_U  =  \langle \psi| U^\dagger (t)\mathcal{O} U(t) |\psi \rangle. 
\end{align*}

While implementing the ideal time-evolution operator using a universal gate set is theoretically possible~\cite{qsim_algo_LCU,qsim_algo_LCU2,qsim_algo_QSP,qsim_algo_quantumwalk,qsim_algo_quantumwalk2,qsim_algo_quantumwalk3,qsim_algo_qubitization,qsim_algo_qwalk}, ideal compilation is generally intractable for many-body systems since realizing such deep circuits is severely constrained in practice by limited hardware connectivity and non-ideal gate fidelities. 

To estimate $\langle \mathcal{O}(t)\rangle_U$, one needs a quantum algorithm to approximate $U(t)$. A standard and practical approach is Trotterization, which approximates $U(t)$ as a sequential product of the unitary evolutions of individual Hamiltonian components $H_i$. For instance, the simplest product formula can be written as
\begin{align*}
 V_2(t) &= \prod_i e^{-i H_i t} = U(t) + E_2(t),
\end{align*}
where $E_2(t) = \mathcal{O}(t^2)$ is the Trotter error originating from the non-commutative properties of unitary operators. In general, we denote a Trotter formula by $V_\alpha(t)$ when it incurs a Trotter error $E_\alpha(t)$ characterized by a leading power $\mathcal{O}(t^{\alpha})$. The specific sequence of the product determines the Trotter errors scaling~\cite{Suzuki1, Suzuki2,qsim_appl2,qsim_appl,PF_second,PF_ruth2,PF_ruth,PF_random2,PF_random,PF_chin}. 

Consequently, for a given initial state $|\psi\rangle$, the expectation value obtained from the Trotterized circuit $V_\alpha(t)$ is
\begin{align*}
\langle \mathcal{O}(t)\rangle_{V_\alpha} &= \langle \psi | V_\alpha^\dagger(t) \mathcal{O} V_\alpha(t) |\psi\rangle, \\
&= \langle \mathcal{O}(t)\rangle_U +  \varepsilon_\alpha(t),
\end{align*}
where the observable Trotter error $\varepsilon_\alpha(t)$ is explicitly written as
\begin{align*}
\varepsilon_\alpha(t) &= \langle \psi | E_\alpha^\dagger(t) \mathcal{O} U(t)+h.c.| \psi \rangle  + \langle \psi | E_\alpha^\dagger(t) \mathcal{O} E_\alpha(t) | \psi \rangle.
\end{align*}

\subsection{Symmetries of the Hamiltonian}
Let $\mathcal{S}$ be a symmetry group of the target Hamiltonian $H$. A symmetry operation $g \in \mathcal{S}$ leaves the Hamiltonian invariant, meaning
\begin{align*}
g H g^\dagger = H, \quad \Leftrightarrow \quad [g, H] = 0, \quad \forall g \in \mathcal{S}. 
\end{align*}
Examples of such symmetry operations include lattice translations, spatial reflections in crystalline systems, global spin rotations in the $XY$ or Heisenberg models, and the anti-unitary time-reversal symmetry. 

Because the symmetry operator commutes with the Hamiltonian, it also commutes with the ideal time-evolution operator, $[g, e^{-i H t}]=0$. Consequently, the ideal time evolution of a state within a specific {irreducible representation (irrep)} subspace remains strictly within that same subspace.

However, the Trotterized time-evolution operator $V_\alpha(t)$ often fails to respect these symmetries. In practice, compiling the Trotter formula into hardware-native gates under restricted connectivity and a limited native gate set can introduce Trotter errors that do not commute with the symmetry operations, so that symmetry is already broken at the circuit level.
Although the total Hamiltonian $H$ is symmetric, its decomposition into non-commuting local terms can break the symmetry of the system. This leads to
\begin{align*}
    g V_\alpha(t) g^\dagger = g U(t) g^\dagger + g E_\alpha(t) g^\dagger \neq V_\alpha(t).
\end{align*}
Since the ideal evolution is symmetric $g U(t) g^\dagger = U(t)$, this symmetry violation strictly originates from the Trotter error, meaning $[g, E_\alpha(t)] \neq 0$. As a result, the wave-function obtained from the Trotterized circuit escapes its initial irrep subspace, resulting in unphysical dynamics that break the selection rules. 

Enforcing these symmetry transformations directly within a time-evolution circuit to prevent such violations can be highly resource-intensive or even impossible. For instance, non-local operations such as spatial reflections or rotation symmetries require long-range qubit connectivity that is lacking in near-term hardware. Furthermore, anti-unitary operations like time-reversal cannot be implemented using standard unitary quantum gates. Rather than strictly enforcing these symmetries in-line at every time step, we can circumvent these hardware limitations by shifting the symmetry operations either to the initial state preparation or to discrete intervals between Trotter circuit layers, followed by a classical post-processing step. Thus, to restore the broken symmetries and mitigate the symmetry-violating contributions of the Trotter error, we propose a post-processing protocol that systematically exploits the symmetries of the target Hamiltonian.

\section{Post-processed symmetry restoration protocol}
\label{protocol}
In this section, we propose a post-processing protocol that mitigates Trotter errors leveraging the symmetries of the target Hamiltonian. The core idea is to evaluate an ensemble of symmetry-transformed quantum simulations and average their results. This procedure effectively symmetrizes the Trotter error and projects out the symmetry-violating unphysical contributions from the expectation values.

To construct the symmetrized Trotter error, we first aim to evaluate the expectation value under an effectively transformed Trotter formula
\begin{align*}
V_{\alpha, g}(t) \equiv g^\dagger V_\alpha(t) g,
\end{align*}
where $g$ is a symmetry operation of the symmetry group $\mathcal{S}$. As discussed in the previous section, directly compiling $V_{\alpha, g}(t)$ is practically challenging. However, we can passively realize this transformation by evaluating the observable with a symmetrically transformed initial state $|\psi_g\rangle = g|\psi\rangle$, {whose preparation requires only a $\mathcal{O}(1)$ quantum circuit depth.} 

Here, we assume that the physical observable $\mathcal{O}$ respects the symmetry, i.e., $[\mathcal{O}, g] = 0$. This condition is satisfied for many standard observables, such as magnetizations and currents in translationally invariant condensed matter systems, or gauge-invariant operators in lattice gauge theories where $g$ represents a local gauge transformation. The more general case $[\mathcal{O}, g] \neq 0$, and how our protocol can be extended to treat it, will be discussed in the discussion section.
%\schoi{what about discuss it this way? discuss about general case where we don't require $[\mathcal{O}, g] = 0$. and then discuss how it can be simplified when $[\mathcal{O}, g] = 0$}

Under this assumption, a quantum simulation using the transformed initial state yields the expectation value
\begin{equation}
\begin{aligned}
\langle \mathcal{O}(t)\rangle_{V_g} &= \langle \psi_g | V_\alpha^\dagger(t) \mathcal{O}  V_\alpha(t) |\psi_g\rangle, \\
&= \langle \psi | \left(g^\dagger V_\alpha^\dagger(t)g\right) \mathcal{O} \left( g^\dagger V_\alpha(t) g\right) |\psi\rangle, \\
&= \langle \psi | V_{\alpha, g}^\dagger(t) \mathcal{O} V_{\alpha, g}(t) |\psi\rangle.
\end{aligned}
\end{equation}
Note that while we described this using initial state preparation, an equivalent mathematical outcome is achieved by inserting $g$ operations at discrete intervals within the circuit, which will be demonstrated later.

By substituting the relation $V_\alpha(t) = U(t) + E_\alpha(t)$ and utilizing the fact that the ideal evolution commutes with the symmetry $[g, U(t)] = 0$, the effectively transformed expectation value decomposes as
\begin{equation}
\begin{aligned}
\langle \mathcal{O}(t)\rangle_{V_g} &= \langle \mathcal{O}(t)\rangle_U + \varepsilon^{(1)}_{\alpha,g}(t) + \varepsilon^{(2)}_{\alpha,g}(t),\\
\varepsilon^{(1)}_{\alpha,g}(t) &= \langle \psi | U^\dagger(t) \mathcal{O} \left(g^\dagger E_{\alpha}(t) g \right) |\psi\rangle + c.c., \\
\varepsilon^{(2)}_{\alpha,g}(t) &= \langle \psi | \left( g^\dagger E^\dagger_{\alpha}(t) g \right) \mathcal{O} \left( g^\dagger E_{\alpha}(t) g \right) |\psi\rangle,
\label{error}
\end{aligned}
\end{equation}
where the Trotter errors exhibit error scaling $\varepsilon^{(1)}_{\alpha,g}(t) = \mathcal{O}(t^{\alpha})$ and $\varepsilon^{(2)}_{\alpha,g}(t) = \mathcal{O}(t^{2\alpha})$. 

Based on this decomposition, we systematically mitigate the symmetry-violating Trotter errors by simulating the circuits for a set of generators (or a well-chosen subset) of $\mathcal{S}$, and averaging the obtained measurement outcomes
\begin{align}
\langle \mathcal{O}(t)\rangle_{\bar{\mathcal{S}}} = \frac{1}{|\mathcal{S}|} \sum_{g\in \mathcal{S}} \langle \mathcal{O}(t)\rangle_{V_g}.
\end{align}
We argue that the averaging process strictly eliminates the symmetry-violating contributions. 

To show that, let us define the symmetrized Trotter error operator
\begin{align}
\bar{E}_\alpha(t) = \frac{1}{|\mathcal{S}|} \sum_{g\in \mathcal{S}} g^\dagger E_\alpha(t) g.
 \end{align} 
By the rearrangement theorem in group theory, the averaged operator is invariant under conjugation by any $h \in \mathcal{S}$, implying that $[\bar{E}_\alpha(t),h] = 0$ for all $h \in \mathcal{S}$. To rigorously see the error cancellation even in the presence of degeneracies, we decompose the initial state into the basis of the group's irreps
\begin{align}
|\psi\rangle = \sum_{\lambda, n, i} a_{\lambda, n, i} |\varphi_{\lambda, n, i} \rangle,
\end{align}
where $\lambda$ labels the irrep, $i$ is the degeneracy index within the irrep, and $n$ accounts for multiple occurrences of the same irrep (multiplicity). A symmetry operation $g \in \mathcal{S}$ acts on this basis as $g |\varphi_{\lambda, n, i} \rangle = \sum_j [D^\lambda(g)]_{ji} |\varphi_{\lambda, n, j} \rangle$, where $D^\lambda(g)$ is the representation matrix {of symmetry operation $g$}. 

The matrix elements of the symmetrized Trotter error operator are strictly constrained by the Great Orthogonality Theorem. For the leading-order contribution, we have:
\begin{equation}
\begin{aligned}
\langle \varphi_{\lambda, n, i} | \bar{E}_\alpha(t) | \varphi_{\lambda', n', i'} \rangle &= \frac{1}{|\mathcal{S}|} \sum_{g \in \mathcal{S}} \langle \varphi_{\lambda, n, i} | g^\dagger E_\alpha(t) g | \varphi_{\lambda', n', i'} \rangle, \\
&= \frac{1}{d_\lambda} \delta_{\lambda \lambda'} \delta_{ii'} \sum_{j} \langle \varphi_{\lambda, n, j} | E_\alpha(t) | \varphi_{\lambda, n', j} \rangle.
\end{aligned}
\end{equation}
Similarly, the averaging procedure systematically eliminates unphysical inter-irrep transitions from the higher-order error term $\varepsilon^{(2)}_{\alpha,g}(t)$. By utilizing the group average over the transformed quadratic operator $g^\dagger E^\dagger_\alpha(t) \mathcal{O} E_\alpha(t) g$, the Great Orthogonality Theorem yields:
\begin{equation}
\begin{aligned}
&\frac{1}{|\mathcal{S}|} \sum_{g \in \mathcal{S}} \langle \varphi_{\lambda, n, i} | g^\dagger E^\dagger_\alpha(t) \mathcal{O} E_\alpha(t) g | \varphi_{\lambda', n', i'} \rangle \\
=& \frac{1}{d_\lambda} \delta_{\lambda \lambda'} \delta_{ii'} \sum_{j} \langle \varphi_{\lambda, n, j} | E^\dagger_\alpha(t) \mathcal{O} E_\alpha(t) | \varphi_{\lambda, n', j} \rangle,
\end{aligned}
\end{equation}
where we have used the fact that $[D^\lambda(g), \mathcal{O}] = 0$. This shows that both the linear and quadratic components of the Trotter error are effectively block-diagonalized with respect to the irrep labels and act as the identity within each degenerate subspace (Schur's Lemma). Consequently, the total mitigated error strictly reduces to block-diagonal intra-irrep dynamics:
\begin{widetext}
\begin{equation}
\begin{aligned}
\langle \mathcal{O} (t)  \rangle_{\bar{\mathcal{S}}} - \langle \mathcal{O} (t)  \rangle_{U} &= \sum_{\lambda} \sum_{n, n'} \mathcal{A}_{\lambda, n, n'} \frac{1}{d_\lambda} \sum_{j} \Big[ \langle \varphi_{\lambda, n, j} | U^\dagger(t) \mathcal{O} E_\alpha(t) |\varphi_{\lambda, n', j} \rangle + c.c. \Big] \\
&\quad + \sum_{\lambda} \sum_{n, n'} \mathcal{A}_{\lambda, n, n'} \frac{1}{d_\lambda} \sum_{j} \langle \varphi_{\lambda, n, j} | E^\dagger_\alpha(t) \mathcal{O} E_\alpha(t) |\varphi_{\lambda, n', j} \rangle,
\end{aligned}
\end{equation}
\end{widetext}
where $\mathcal{A}_{\lambda, n, n'} = \sum_i a^*_{\lambda, n, i} a_{\lambda, n', i}$ is the overlap matrix within the irrep subspace. Because the dominant symmetry-violating components of both the $\mathcal{O}(t^\alpha)$ and $\mathcal{O}(t^{2\alpha})$ errors are projected out, the overall observable error scaling is improved, ultimately bounded only by the remaining symmetry-preserving physical dynamics.

Conversely, without symmetry mitigation, the raw Trotter error inevitably contains terms proportional to $a^*_{\lambda', n', i'} a_{\lambda, n, i} \langle \varphi_{\lambda', n', i'} | U^\dagger(t) \mathcal{O} E_\alpha(t) |\varphi_{\lambda, n, i} \rangle$, representing unphysical inter-subspace transitions ($\lambda \neq \lambda'$) induced by the symmetry-violating Trotter components. By applying our post-processing protocol, we successfully filter out these unphysical contributions, ensuring robust quantum simulations.

{
The initial-state symmetrization protocol can be naturally extended to an interleaved circuit approach in which symmetry transformations are inserted between discrete Trotter layers. While previous studies have explored interleaving symmetry operations to modify the coherent error structure or employed measurement-based post-selection to discard symmetry-violating states~\cite{qsim_algo_sym, gauge_sym_verification}, these approaches strictly require deep coherent circuits or substantial hardware overhead. By contrast, our framework leverages classical post-processing: we trade coherent circuit depth for classical measurement overhead by evaluating an ensemble of shallow, symmetry-interleaved circuits and averaging their outcomes to systematically twirl away the error.

Let $V_\alpha(\tau)$ be a single Trotter step with a time slice $\tau=t/N$, such that the total unmitigated evolution is $V_\alpha(t) = [V_\alpha(\tau)]^N$. For a sequence of symmetry operations drawn from the symmetry group, $\boldsymbol{g}=(g_1,\dots,g_N)\in \mathcal{S}^N$, we define the effectively interleaved, symmetry-transformed evolution operator as
\begin{equation}
V_{\alpha, \boldsymbol{g}}(t)
=
\prod_{k=N}^{1}
\left(
g_k^\dagger V_\alpha(\tau) g_k
\right)
=
g_N^\dagger V_\alpha(\tau) g_N
\cdots
g_1^\dagger V_\alpha(\tau) g_1.
\label{eq:interleaved_circuit}
\end{equation}
In practice, implementing this operator only requires inserting the composite transformation $g_{k}^\dagger g_{k+1}$ between adjacent Trotter layers. Because each symmetry operation commutes with the ideal short-time evolution ($[g_k, U(\tau)]=0$), the ideal dynamics remain entirely unaffected, $g_k^\dagger U(\tau) g_k = U(\tau)$. However, the Trotter error $E_\alpha(\tau)$ transforms non-trivially. 

Assuming the observable is symmetry-invariant, $[\mathcal{O}, g]=0$, the expectation value associated with a specific symmetry sequence $\boldsymbol{g}$ is
\begin{equation}
\langle \mathcal{O}(t)\rangle_{V_{\boldsymbol{g}}}
=
\langle \psi |
V_{\alpha, \boldsymbol{g}}^\dagger(t)\,
\mathcal{O}\,
V_{\alpha, \boldsymbol{g}}(t)
|\psi\rangle .
\label{eq:Ovgbold}
\end{equation}
We then construct our mitigated outcome by averaging over an ensemble $\mathcal{G}_{\mathrm{ens}} \subseteq \mathcal{S}^N$ of these sequences,
\begin{equation}
\langle \mathcal{O}(t)\rangle_{\mathrm{inter}}
=
\frac{1}{|\mathcal{G}_{\mathrm{ens}}|}
\sum_{\boldsymbol{g}\in \mathcal{G}_{\mathrm{ens}}}
\langle \mathcal{O}(t)\rangle_{V_{\boldsymbol{g}}}.
\label{eq:interleaved_average}
\end{equation}

To explicitly demonstrate the mitigation mechanism, we substitute $V_\alpha(\tau) = U(\tau) + E_\alpha(\tau)$ into Eq.~\eqref{eq:interleaved_circuit}. Expanding the interleaved circuit to leading order in the Trotter error yields
\begin{equation}
V_{\alpha, \boldsymbol{g}}(t)= U(t)+\sum_{k=1}^{N}U\big((N-k)\tau\big)\,
\big[ g_k^\dagger E_\alpha(\tau) g_k \big]\,
U\big((k-1)\tau\big)
+ \mathcal{O}(E_\alpha^2).
\label{eq:interleaved_expansion}
\end{equation}
When the ensemble average in Eq.~\eqref{eq:interleaved_average} is evaluated over the symmetry group, the local error contribution at each step is independently projected onto its symmetrized form,
\begin{equation}
\bar{E}_\alpha(\tau)
=
\frac{1}{|\mathcal{S}|}
\sum_{g\in\mathcal{S}}
g^\dagger E_\alpha(\tau) g.
\label{eq:symmetrized_error_layer}
\end{equation}
By construction, this averaged error operator commutes with all symmetry elements ($[h, \bar{E}_\alpha(\tau)] = 0$ for all $h \in \mathcal{S}$). Thus, identical to the initial-state protocol, the group average block-diagonalizes the error operator with respect to the irreducible representations of $\mathcal{S}$. 

This highlights a key advantage: rather than relying on deep quantum circuits to directly suppress errors, we  apply symmetry twirling to the local Trotter errors layer-by-layer. This systematically projects out leakage into unphysical subspaces during post-processing, providing a scalable and depth-preserving method for near-term hardware.
}

\section{Numerical demonstrations}
\label{numerics}
\subsection{Crystal symmetries in condensed matter systems}
In condensed matter physics, intrinsic symmetries play a pivotal role in characterizing physical phases. However, strictly preserving these symmetries during a digital quantum simulation is notoriously difficult. For instance, spatial point-group symmetries inherently require non-local operations. In a 2D fermionic system encoded via the Jordan-Wigner transformation, a simple $C_4$ rotational symmetry becomes a highly complex, non-local string operation demanding prohibitive long-range qubit connectivity. Furthermore, time-reversal symmetry requires anti-unitary operations (e.g., complex conjugation), which are fundamentally impossible to implement using standard unitary quantum gates during the time evolution. 

Our post-processing protocol bypasses these hardware and algorithmic limitations. By shifting these challenging symmetry operations whether they are non-local spatial transformations or anti-unitary time reversals entirely to the initial state preparation, we can efficiently restore the broken symmetries without adding any burden to the circuit depth.

To demonstrate this, we apply our protocol to the 1D $XY$ model with spatial reflection symmetry. We consider a 7-site open chain governed by the Hamiltonian $H = H_{\text{odd}} + H_{\text{even}}$, where
\begin{align*}
H_{\text{odd}} &= \sum_{k=1}^{3} \left(S^x_{2k-1} S^x_{2k} + S^y_{2k-1} S^y_{2k} \right),\\
H_{\text{even}} &= \sum_{k=1}^{3} \left(S^x_{2k} S^x_{2k+1} + S^y_{2k} S^y_{2k+1} \right).
\end{align*}
We focus on the dynamics of the total magnetization in the $x$-direction, $\mathcal{O} = \sum_{i=1}^{7} S^x_i$, starting from a product state
\begin{align*}
|\psi \rangle =\frac{1}{2^{7/2}}\bigotimes_{s = 1}^{7} 
\begin{pmatrix}
1\\
e^{-i\frac{2\pi}{7}s }
\end{pmatrix}.
\end{align*}

To simulate the time evolution, we employ the first-order Trotter product formula $V_2(t) = e^{-i H_{\text{odd}} t} e^{-i H_{\text{even}} t}$. Under the spatial site reflection operator $\mathcal{R}$, the sub-Hamiltonians map to each other $\mathcal{R} H_{\text{odd}} \mathcal{R}^\dagger = H_{\text{even}}$. Consequently, the total Hamiltonian and the observable are invariant under reflection ($[\mathcal{R}, H] = 0$, $[\mathcal{R}, \mathcal{O}] = 0$). However, the Trotterized time-evolution operator is explicitly not invariant
\begin{align*}
\mathcal{R} V_2(t) \mathcal{R}^\dagger = e^{-i H_{\text{even}} t} e^{-i H_{\text{odd}} t} \neq V_2(t).
\end{align*}
This inconsistency fundamentally breaks the reflection symmetry during the quantum simulation, leading to unphysical symmetry-violating algorithmic errors.

\begin{figure}[t!]
\centering
 \includegraphics[width=1\linewidth]{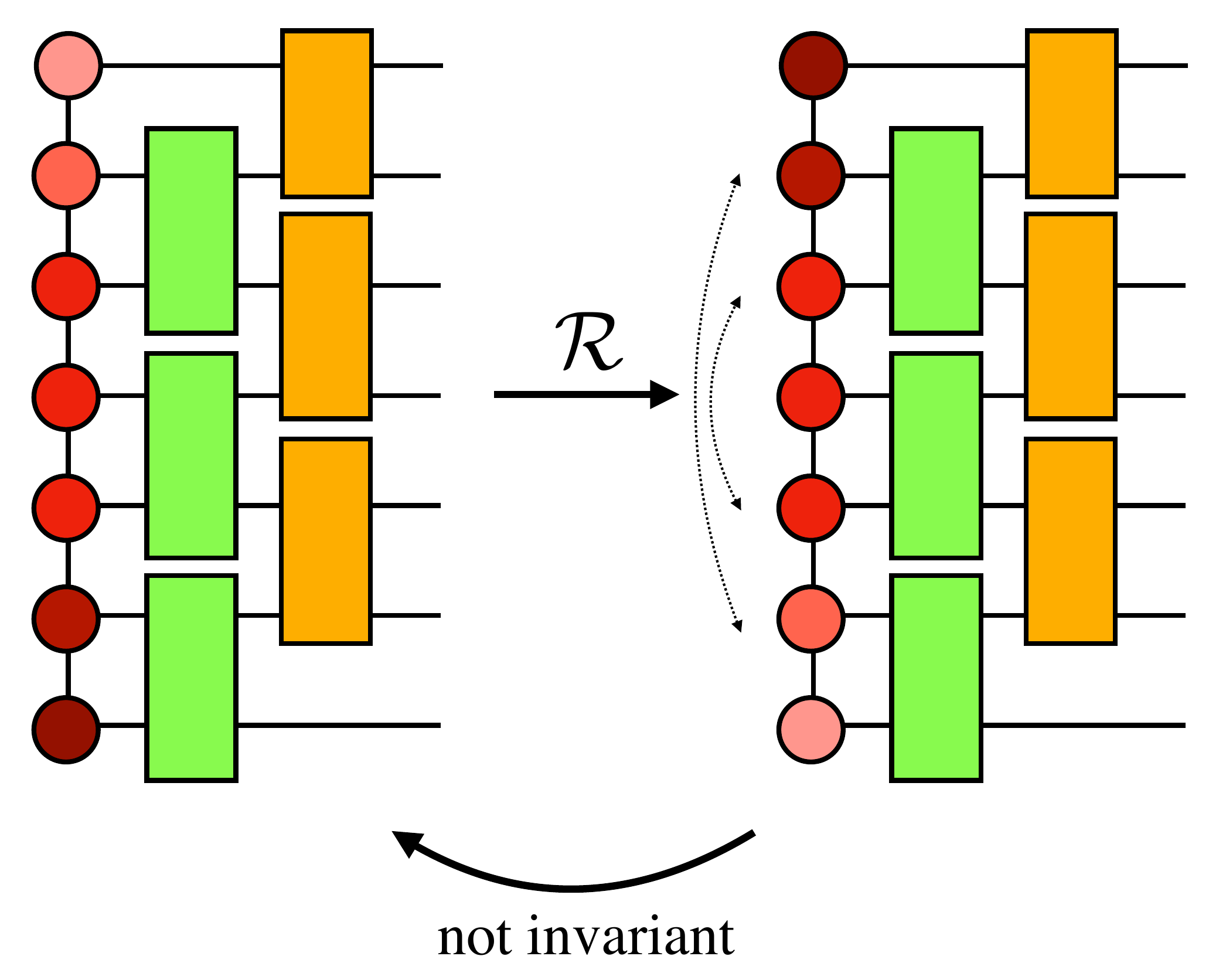}
  \caption{Inconsistency of symmetry operations in quantum circuits. A physical symmetry of the Hamiltonian should be preserved during quantum simulation. However, some symmetries (here, a reflection symmetry $\mathcal{R}$) are explicitly broken when compiling the dynamics via Trotterization. These symmetry-violating contributions manifest as unphysical Trotter errors in the measurement outcomes.}
  \label{Fig1}
\end{figure}

To enforce the reflection symmetry and mitigate the Trotter error without executing the non-local $\mathcal{R}$ operation on the quantum circuit, we prepare a symmetrically reflected initial state
\begin{align*}
|\psi_{\mathcal{R}}\rangle \equiv \mathcal{R}|\psi\rangle = \frac{1}{2^{7/2}}\bigotimes_{s = 1}^{7} 
\begin{pmatrix}
1\\
e^{-i\frac{2\pi}{7}(8-s) }
\end{pmatrix}.
\end{align*}
By independently measuring the observable using $|\psi\rangle$ and $|\psi_{\mathcal{R}}\rangle$, and subsequently averaging the results according to our protocol $\langle \mathcal{O}(t)\rangle_{\bar{\mathcal{S}}} = \frac{1}{2} \left( \langle \mathcal{O}(t)\rangle_{V_2, |\psi\rangle} + \langle \mathcal{O}(t)\rangle_{V_2, |\psi_{\mathcal{R}}\rangle} \right)$, we systematically project out the reflection-symmetry-violating errors.

\begin{figure}[t!]
\centering
 \includegraphics[width=1\linewidth]{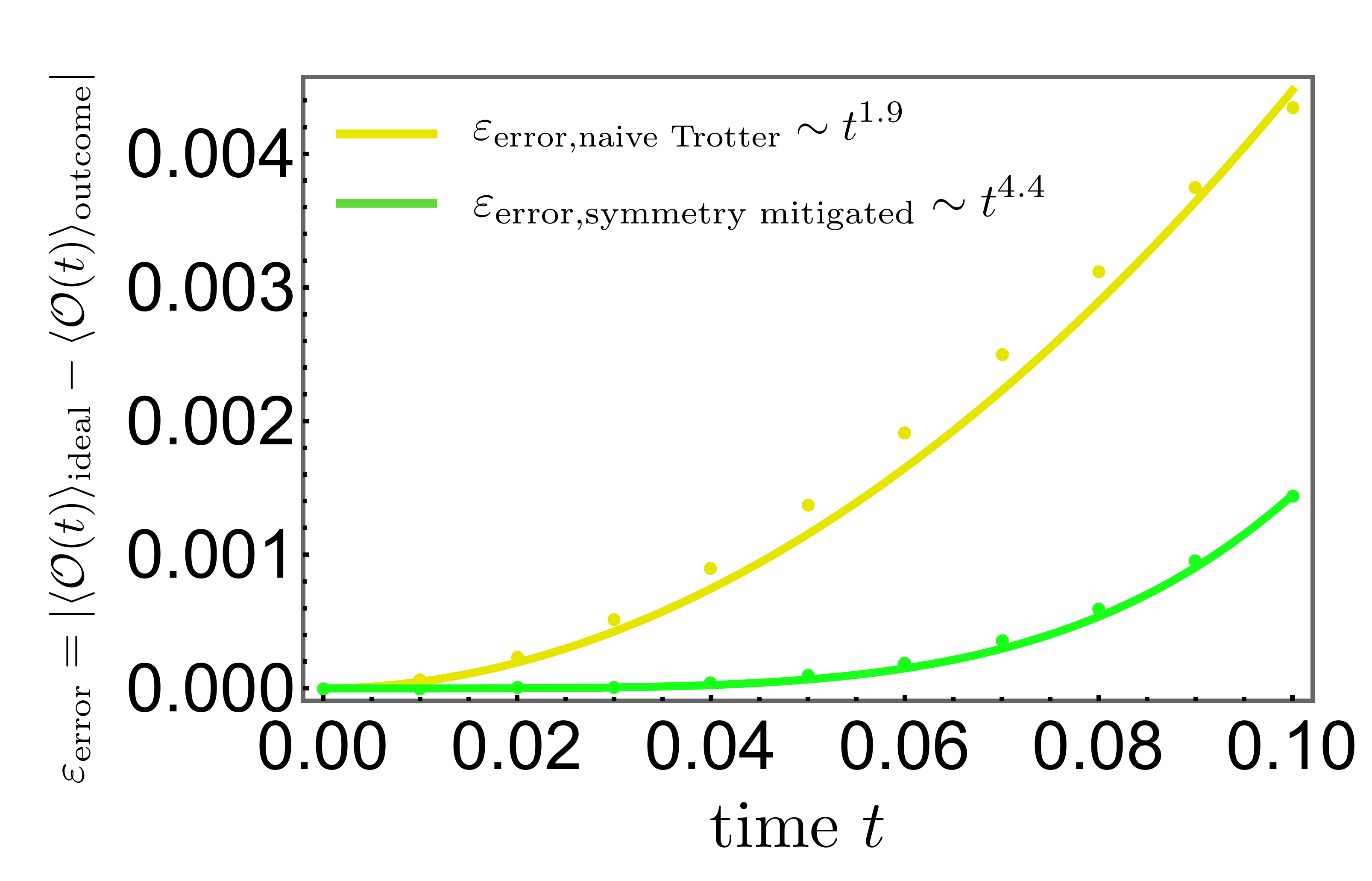}
  \caption{Symmetry-mitigated results for the seven-site one-dimensional $XY$ model. Yellow markers represent the expectation values obtained by the naive first-order Trotter simulation ($V_2$), and green markers correspond to the symmetry-mitigated results using our post-processing protocol. Solid lines indicate polynomial fits. The naive Trotter error scales as $\mathcal{O}(t^{1.9})$, while the symmetry-mitigated error is drastically suppressed to $\mathcal{O}(t^{4.4})$.}
  \label{Fig2}
\end{figure}

Our numerical results (see Fig.~\ref{Fig2}) shows mitigated result by our proposal. The naive first-order Trotter simulation exhibits a leading Trotter error scaling of approximately $\mathcal{O}(t^{1.9})$, consistent with the expected $\mathcal{O}(t^2)$ theoretical bound for the $V_2(t)$ formula. In contrast, applying the symmetry-mitigated protocol suppresses the algorithmic error, yielding an improved scaling of $\mathcal{O}(t^{4.4})$. This scaling enhancement confirms that the dominant symmetry-violating components of the Trotter error are systematically filtered out through our post-processing scheme.

\subsection{Gauge invariance in lattice gauge theory}
{
In a lattice gauge theory (LGT), local gauge invariance is a fundamental governing principle~\cite{LGT_review, LGT2}. For a given LGT Hamiltonian $H$, there exists a set of local Gauss's law generators $G_i$ on every site $i$ such that $[H, G_i] = 0$, and any physical state $|\psi\rangle$ must satisfy $G_i |\psi\rangle = 0$ for all $i$. Consequently, the ideal time evolution $e^{-i H t}$ restricts the dynamics strictly within the physical subspace.

However, in the quantum simulation of LGTs, compiling the Hamiltonian into a Trotterized quantum circuit inevitably breaks this local gauge invariance. A local gauge transformation, characterized by gauge angles $\vec{a}$, is given by $U_{\vec{a}} = \prod_i e^{-i G_i a_i } = e^{-i\vec{G}\cdot\vec{a}}$. This operation acts nontrivially on the Trotterized circuit, meaning $[U_{\vec{a}}, V_\alpha(t)] \neq 0$. As a result, the Trotter error splits into a gauge-invariant component and a gauge-violating component. This gauge-violating component is directly responsible for the unphysical gauge violations observed in LGT simulations. While applying a large energy penalty to suppress such violations is a common approach~\cite{gauge_penalty, gauge_penalty2,gauge_protection}, we instead apply our symmetry post-processing protocol to actively twirl away the unphysical errors.

Before applying the protocol, we must note that applying the gauge transformation solely to the initial state, as done in the spatial symmetry case, is fundamentally ineffective here. Because the physical initial state is intrinsically symmetric  $U_{\vec{a}}|\psi\rangle = |\psi\rangle$) and any physical observable must be gauge-invariant $[\mathcal{O}, U_{\vec{a}}] = 0$), the gauge-transformed expectation value is mathematically identical to the unmitigated one.

To circumvent this and successfully mitigate gauge violations, we must invoke the interleaved protocol introduced in Section~\ref{protocol}. Since the $U(1)$ gauge transformations form an abelian group, the intermediate operations between Trotter layers reduce to $U_{\vec{a}_k}^\dagger U_{\vec{a}_{k+1}} = U_{\vec{a}_{k+1} - \vec{a}_k}$. Because the gauge angles are arbitrary, we can simply define a set of relative random gauge angles $\vec{a}_s$ to be inserted between the layers. The interleaved circuit is thus explicitly compiled as:
\begin{align}
V_{N,\alpha, \{\vec{a}_s\}} (t) &= V_\alpha\left(\frac{t}{N}\right)\prod_{s=1}^{N-1} \left(e^{-i \vec{G}\cdot \vec{a}_s} V_\alpha\left(\frac{t}{N}\right)\right).
\end{align}
By simulating this interwoven circuit for an ensemble of random gauge choices $\{\vec{a}_s\}$ and averaging the corresponding measurement outcomes, we effectively perform an in-circuit twirling over the $U(1)$ gauge group:
\begin{align}
\langle \mathcal{O}(t) \rangle_{\mathrm{inter}} = \int \prod_{s=1}^{N-1} d\vec{a}_s \, \langle \psi| V_{N,\alpha,\{\vec{a}_s\}}^\dagger(t) \mathcal{O} V_{N,\alpha,\{\vec{a}_s\}}(t)| \psi \rangle.
\end{align}
In practice, this continuous integral is approximated by a finite discrete sum over sampled gauge configurations. While one could sample these angles purely at random, the gauge angles can be strategically optimized. By leveraging the group-theoretic transformation properties of the gauge-violating error terms, a carefully chosen minimal set of discrete angles is sufficient to exactly evaluate the integral. This efficient averaging effectively projects out the gauge-violating Trotter errors layer-by-layer.
}

To demonstrate this specifically, we apply the protocol to the 1D quantum electrodynamics with a fermionic matter field, known as the 1D Schwinger model (see the Appendix.~\ref{schwinger_review} for a brief review). The Hamiltonian is $H_S = H_E + H_m + H_F$, where
\begin{align*}
H_E &= c \sum_i E_{i+\frac{1}{2}}^2,\\
H_m &= m \sum_i (-1)^i \psi^\dagger_i \psi_i,\\
H_F &= g \sum_i  \left(\psi^\dagger_{i+1} U_{i+\frac{1}{2}}\psi_{i} + h.c.\right).
\end{align*}
Here, the fields and gauge variables $\{ E_{i+\frac{1}{2}}, U_{i+\frac{1}{2}}\} $ reside on links $(i,i+1)\doteq i+\frac{1}{2}$, and the fermionic matter fields $\psi_i $ reside on sites $i$. 
The associated Gauss's law is
\begin{align}
 G_i  = -E_{i-\frac{1}{2}} + E_{i+\frac{1}{2}} + \psi^\dagger_i \psi_i -\frac{1-(-1)^i}{2}.
 \label{gausslaw}
\end{align}

In the following, we set the coupling constants to $(c,m,g)=(-2,2,-8)$ for simplicity of presentation, however, we emphasize that our symmetry post-processing protocol is generally applicable regardless of the specific choice of system parameters. 

\begin{figure}[t!]
\centering
 \includegraphics[width=1\linewidth]{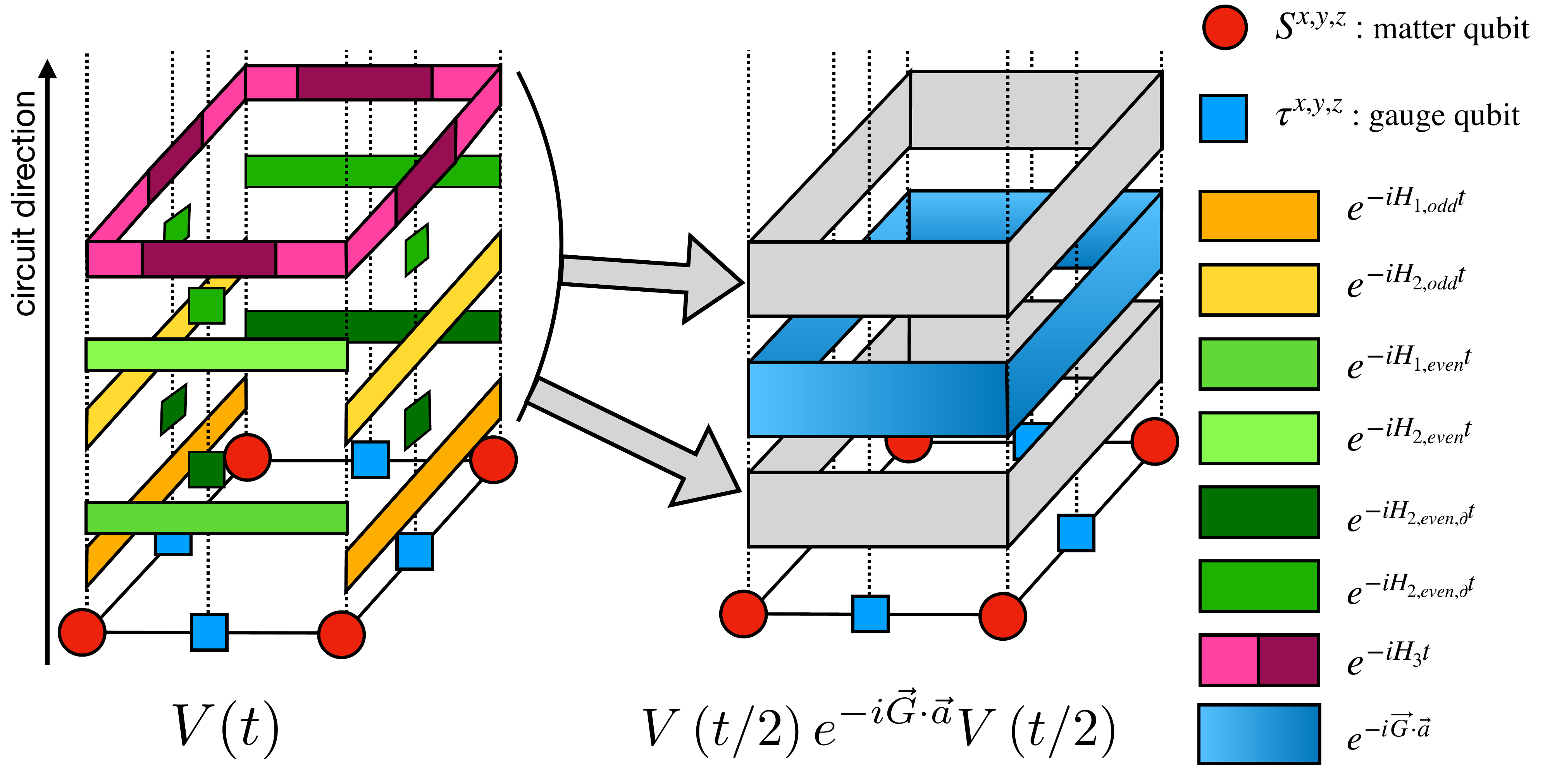}
  \caption{Trotterized circuit for the 1D Schwinger model. Red circles and blue squares represent qubit-encoded matter and gauge degrees of freedom, respectively. The intermediate gauge transformation $e^{-i \vec{G} \cdot \vec{a}}$ is inserted between two $V_2(t/2)$ Trotter layers to mitigate covariant errors.}
  \label{Fig3}
\end{figure}

In standard quantum simulations, the gauge degree of freedom is truncated to a finite dimensional qudit. For simplicity, we focus on a quantum link model using qubits ($E_i \in \{-1/2, 1/2\}$) and map the fermions to spins via the Jordan-Wigner transformation. The Hamiltonian maps to Pauli operators ($S$ for matter, $\tau$ for gauge degrees of freedom)
{\small
\begin{align*}
H_E &= \sum_i \tau^z_{i+\frac{1}{2}},\\
H_m &= \sum_i (-1)^i  S^z_i ,\\
H_F &= \sum_i \Big[ \tau_{i+\frac{1}{2}}^x (S^x_{i+1}S^x_i + S^y_{i+1} S^y_{i} ) + \tau_{i+\frac{1}{2}}^y(S^y_{i+1} S^x_i - S^x_{i+1}S^y_{i} ) \Big],
\end{align*}
}and the mapped Gauss's law is $G_i = \frac{1}{2} \left(\tau^z_{i-1/2} - \tau^z_{i+1/2} + S^z_i +(-1)^i\right)$. We simulate an 8-qubit system under periodic boundary conditions using the $n=2$ layered Trotter formula,
\begin{align}
V_{n=2, \alpha=2, \vec{a}} = V_2\left(\frac{t}{2} \right) e^{-i \vec{G}\cdot \vec{a}} V_2\left(\frac{t}{2} \right).
\end{align}
The explicit circuit compilation is illustrated in Fig.~\ref{Fig3}.

The leading $\mathcal{O}(t^2)$ Trotter error is governed by commutators of the Hamiltonian components. Notice that under local gauge transformations, each term in $H_F$ product of two vectors under rotation like $X\otimes X$, $Y\otimes X$ under $Z$ axis rotation, while terms in $H_E$ and $H_m$ are scalars. Consequently, the leading Trotter errors are composed of scalar (gauge-invariant) and rank-2 (gauge-violating) components. Because this gauge-violating error transforms tensorially, it can be systematically canceled out via discrete twirling. For example, a covariant error term transforms under $e^{-i G_{i+1} a_{i+1}}$ as
\begin{widetext}
\begin{equation}
\begin{aligned}
&\tau^x_{i+\frac{1}{2}}\tau^x_{i+\frac{3}{2}}\left( S^x_i  S^z_{i+1} S^y_{i+2} -S^y_i  S^z_{i+1} S^x_{i+2} \right) \\
\to& \Big[ \cos^2 (a_{i+1})\tau^x_{i+\frac{1}{2}}\tau^x_{i+\frac{3}{2}} - \sin(a_{i+1})\cos(a_{i+1}) (\tau^x_{i+\frac{1}{2}}\tau^y_{i+\frac{3}{2}}+\tau^y_{i+\frac{1}{2}}\tau^x_{i+\frac{3}{2}})  - \sin^2(a_{i+1}) \tau^y_{i+\frac{1}{2}}\tau^y_{i+\frac{3}{2}} \Big] \left( S^x_i  S^z_{i+1} S^y_{i+2} -S^y_i  S^z_{i+1} S^x_{i+2} \right).
\end{aligned}
\end{equation}
\end{widetext}

Since it is practically inefficient that to integrate over the full continuous $U(1)$ gauge parameters, we discretize the twirling process. To cancel these rank-2 tensor error components (where cross terms $\sin(a)\cos(a)$ vanish and squared terms average out), we select four specific gauge angle sets
\begin{align*}
\vec{a}_1 &= \left(\frac{\pi}{4},\frac{3\pi}{4},\frac{5\pi}{4},\frac{7\pi}{4}\right), &
\vec{a}_2 &= \left(\frac{3\pi}{4},\frac{5\pi}{4},\frac{7\pi}{4},\frac{\pi}{4}\right),\\
\vec{a}_3 &= \left(\frac{5\pi}{4},\frac{7\pi}{4},\frac{\pi}{4},\frac{3\pi}{4}\right), &
\vec{a}_4 &= \left(\frac{7\pi}{4},\frac{\pi}{4},\frac{3\pi}{4},\frac{5\pi}{4}\right).
\end{align*}

\begin{figure}[t!]
\centering
 \includegraphics[width=1\linewidth]{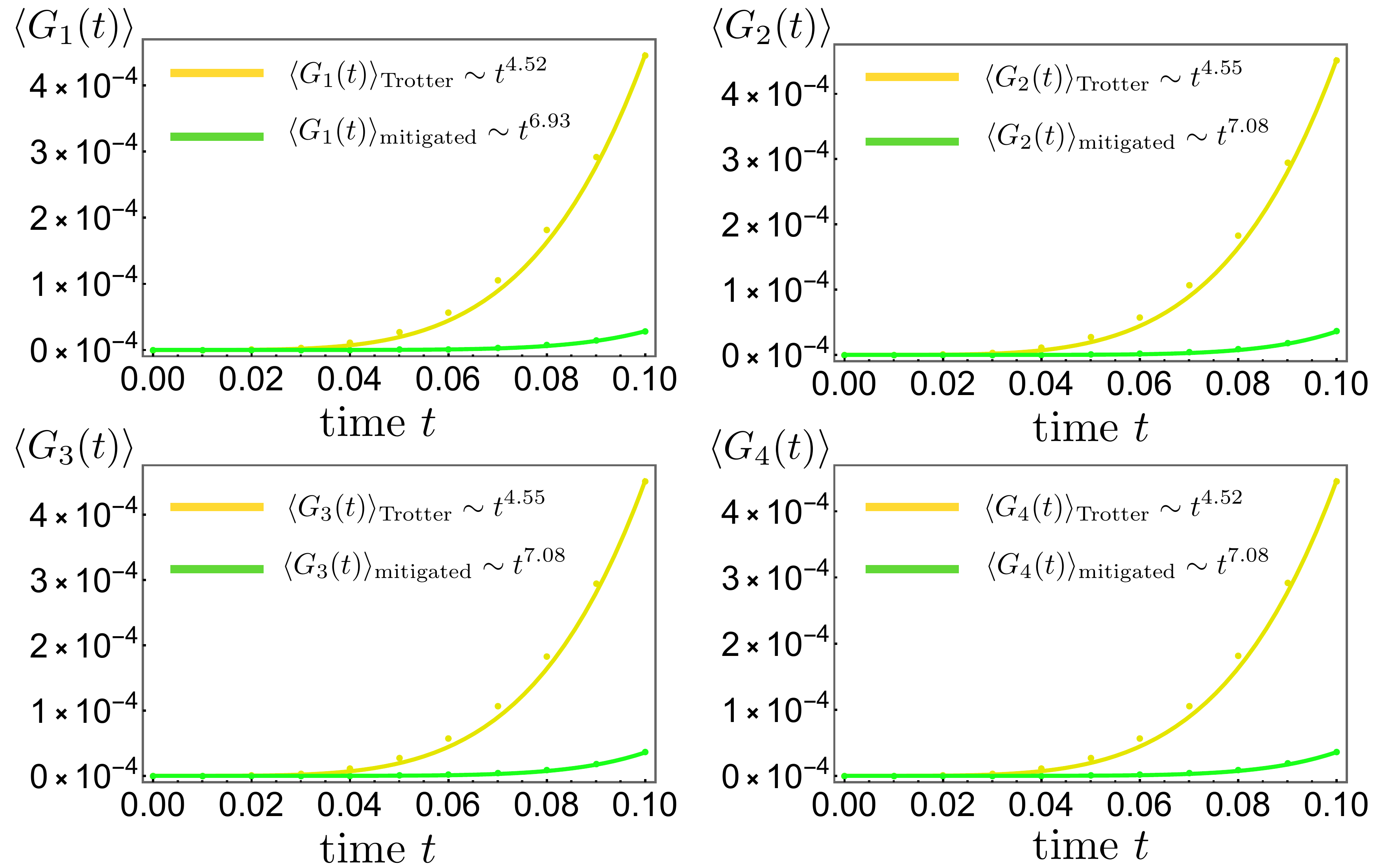}
  \caption{Measurement of gauge violations $\langle G_{i=1,2,3,4}(t) \rangle$ over time in the 1D Schwinger model. The results demonstrate that our intermediate gauge-twirling protocol systematically suppresses the unphysical gauge violations compared to the unmitigated Trotter circuit.}
  \label{Fig4}
\end{figure}

We execute the circuits for these four gauge choices and average the results. To quantify the unphysical algorithmic error, we measure the gauge violation $\langle G_i(t) \rangle$ for all $i$, which remains zero under ideal time evolution. As shown in Fig.~\ref{Fig4}, applying the post-processing gauge mitigation protocol systematically reduces the gauge violations accumulated over the Trotter steps, verifying the suppression of covariant Trotter errors.
We observe that the gauge violation in the naive Trotter circuit scales as $\mathcal{O}(t^{4.5})$, whereas with gauge twirling it is improved to $\mathcal{O}(t^{6.9\text{--}7.1})$, demonstrating a substantial reduction of unphysical gauge violations.

\section{Discussion and Conclusion}
\label{discussion}
In this work, we proposed a symmetry-based post-processing protocol to mitigate Trotter errors in quantum simulations. Symmetries, which dictate the strict selection rules of ideal dynamics, are inevitably broken when the time-evolution operator is compiled into non-commuting Trotterized operators. By leveraging the invariance of physical observables and systematically averaging over an ensemble of symmetry-transformed configurations either through initial state preparation or {layer-by-layer interleaved} twirling,  { both of which require only $\mathcal{O}(1)$ additional circuit depth}, we demonstrated that symmetry-violating Trotter errors can be rigorously projected out. 

A practical advantage of our approach is its minimal overhead in terms of quantum resources. Unlike standard error mitigation techniques that often require deep entangling circuits or ancilla qubits, our protocol offloads the complexity of non-local spatial symmetries and anti-unitary time-reversal symmetries entirely to the initial state preparation. For local gauge symmetries, the intermediate twirling operations require only shallow single-qubit rotations {systematically inserted between Trotter layers}. Consequently, this depth-preserving nature makes our framework well-suited for near-term and early fault-tolerant quantum devices.

Furthermore, the proposed framework can be readily generalized to systems possessing multiple commuting symmetry groups. For a target Hamiltonian governed by a composite symmetry group $\mathcal{S} = \bigotimes_{i=1}^n \mathcal{S}^{(i)}$, where the individual symmetry groups mutually commute ($[\mathcal{S}^{(i)}, \mathcal{S}^{(j)}] = 0$), one can simultaneously mitigate diverse algorithmic errors. By evaluating the ensemble with composite symmetry operations $g = g_{(1)} \otimes g_{(2)} \otimes \cdots \otimes g_{(n)}$, the protocol allows for the simultaneous restoration of various symmetries such as spatial point-group symmetries, particle number conservation, and time-reversal symmetry, comprehensively restricting the dynamics to the correct physical subspace.

 We also note the case where the observable does not commute with the symmetry, $[\mathcal{O}, g] \neq 0$. In this situation, preparing symmetry-transformed initial states {or naively interleaving transformations} is insufficient to symmetrize the Trotter errors. To extend our protocol to this more general setting, one can instead measure the observable in the symmetry-transformed basis, i.e., measure $\mathcal{O}_g = g^\dagger \mathcal{O} g$ {alongside the state or circuit transformations}. This replacement ensures that the leading-order contribution $\varepsilon^{(1)}_{\alpha,g}$ is governed by the same effective operator structure as in the commuting case, so that symmetry-violating contributions can be projected out by the group average. {However,} the higher-order contribution $\varepsilon^{(2)}_{\alpha,g}$ is, in general, only partially constrained. Mitigating {these residual higher-order errors for non-commuting observables} will be an interesting subject for future work.

Our argument also naturally extends to continuous symmetries ubiquitous in physics. In quantum chemistry, preserving $U(1)$ particle number conservation and $SU(2)$ total spin rotation is critical, yet these symmetries are frequently violated by Trotterized basis transformations. In high-energy physics, applying our protocol to non-Abelian lattice gauge theories (such as $SU(2)$ or $SU(3)$) within gauge-invariant subspaces could pave the way for observing intricate phenomena like quark confinement and string breaking on large-scale quantum computers.

Furthermore, while our numerical demonstration for the 1D Schwinger model utilized a specific subset of discrete gauge angles to mitigate the leading-order gauge violations, this framework can be systematically scaled. For simulations involving multiple Trotter layers or higher-order product formulas, the gauge-violating errors manifest as higher-rank tensor components. By {extending our group-theoretic optimization to find} an optimal set of gauge parameters ${\vec{a}_s}$, one can rigorously project out these higher-order covariant errors. This optimization over the discrete gauge choices provides a highly scalable pathway to achieve even deeper mitigation of gauge violations in complex lattice gauge theory simulations.

We emphasize the non-perturbative nature of our protocol. While our numerical demonstrations focused on the small-$t$ regime to explicitly verify the polynomial scaling improvement of the Trotter error, the symmetry-restoration mechanism itself does not rely on a perturbative expansion. Because the post-processing strictly eliminates inter-irrep contributions, it prevents the wavefunction from leaking into unphysical irrep subspaces at any time scale. By combining our symmetry mitigation protocol with advanced quantum simulation algorithms such as randomized compilation~\cite{qdrift, qSWIFT, qshift}, or higher-order product formulas~\cite{PF_chin,PF_ruth,PF_ruth2,PF_second}, one could reliably simulate long-time quantum dynamics, such as thermalization or many-body localization, where Trotter errors typically compound uncontrollably.

Finally, while this work has primarily focused on algorithmic errors, our symmetry post-processing protocol also simplifies error channels of physical hardware noise. Because the ideal expectation value is invariant under symmetry transformations, averaging the measurement outcomes $\langle \mathcal{O}(t) \rangle_{V_g}$ across various symmetry-rotated bases effectively acts as an in-circuit twirl. Similar to the well-known concepts of Pauli twirling and randomized compiling, this process averages out the off-diagonal coherent components of the physical noise. Consequently, it tailors complex hardware error channels into more manageable stochastic errors. This channel simplification provides an intrinsic dual benefit for near-term quantum simulations, making our protocol highly synergistic with standard quantum error mitigation techniques such as zero-noise extrapolation and probabilistic error cancellation~\cite{PEC1,PEC2,ZNE1,ZNE2}. A brief formalization of this physical error simplification is provided in Appendix~\ref{error_simplification}.

In conclusion, our post-processed symmetry restoration protocol offers a scalable, depth-preserving, and physically intuitive pathway to highly accurate digital quantum simulations of complex many-body systems and gauge theories.

\section*{Acknowledgment}
SC was supported by a KIAS Individual Grant (CG090601) at Korea Institute for Advanced Study and by Quantum Simulator Development Project for Materials Innovation through the National Research Foundation of Korea (NRF) funded by the Korean government (Ministry of Science and ICT(MSIT))(No. NRF- 2023M3K5A1094813).  SL is supported by a KIAS Individual Grant via the Quantum Universe Center (QP104301-6P104301) at Korea Institute for Advanced Study and by Institute of Information \& communications Technology Planning \& Evaluation (IITP) grant funded by the Korea government (MSIT) (No.2022-0-01026).  We used resources of the Center for Advanced Computation at Korea Institute for Advanced Study.

 \bibliography{CTEMref}

\newpage
\appendix

\onecolumngrid
\section{Algebraic formulation of the 1D Schwinger model}
\label{schwinger_review}
The 1D Schwinger model describes quantum electrodynamics (QED) in one spatial dimension. To simulate this continuum theory on a digital quantum computer, the model is discretized onto a spatial lattice using the Kogut-Susskind staggered fermion formulation, and the infinite-dimensional gauge fields are truncated via the Quantum Link Model (QLM) approach. In this appendix, we review the algebraic mapping from the fermionic lattice model to the qubit Hamiltonian used in the main text.

\subsection{Kogut-Susskind Lattice Formulation}
In the staggered formulation, the continuous Dirac spinor is split across alternating lattice sites to avoid the fermion doubling problem. Even sites represent positrons (where an empty site corresponds to a positron), and odd sites represent electrons. We denote the fermionic annihilation and creation operators at site $i$ as $\psi_i$ and $\psi_i^\dagger$, satisfying the standard anti-commutation relations $\{\psi_i, \psi_j^\dagger\} = \delta_{i,j}$.

The $U(1)$ gauge fields reside on the links $(i, i+1) \equiv i+\frac{1}{2}$ connecting adjacent sites. The electric field operator $E_{i+\frac{1}{2}}$ and the parallel transporter (gauge link) operator $U_{i+\frac{1}{2}} = e^{i\theta_{i+\frac{1}{2}}}$ satisfy the commutation relation
\begin{align}
[E_{i+\frac{1}{2}}, U_{j+\frac{1}{2}}] = U_{i+\frac{1}{2}} \delta_{i,j}.
\end{align}
This implies that $U$ acts as a raising operator for the electric field $E U |e\rangle = (e+1) U |e\rangle$. 

The unencoded lattice Hamiltonian is given by $H_S = H_E + H_m + H_F$, where
\begin{align}
H_E &= c \sum_i E_{i+\frac{1}{2}}^2,\\
H_m &= m \sum_i (-1)^i \psi^\dagger_i \psi_i,\\
H_F &= g \sum_i  \left(\psi^\dagger_{i+1} U_{i+\frac{1}{2}}\psi_{i} + h.c.\right).
\end{align}
The physical Hilbert space is strictly constrained by the local Gauss's law, which ensures gauge invariance at every site $i$
\begin{align}
 G_i  = -E_{i-\frac{1}{2}} + E_{i+\frac{1}{2}} + \psi^\dagger_i \psi_i -\frac{1-(-1)^i}{2} = 0.
\end{align}

\subsection{Quantum Link Model and Qubit Encoding}
In standard $U(1)$ lattice gauge theory, $E_{i+\frac{1}{2}}$ possesses an infinite discrete spectrum, $e \in \mathbb{Z}$. For quantum simulation, we truncate this infinite-dimensional space into a finite spin-$S$ system (Quantum Link Model). Taking the extreme limit of $S=1/2$, we encode the gauge degrees of freedom into qubits (denoted by Pauli matrices $\tau$). The mapping is defined as
\begin{align}
E_{i+\frac{1}{2}} &\to \frac{1}{2}\left(1 - \tau^z_{i+\frac{1}{2}}\right), \\
U_{i+\frac{1}{2}} &\to \tau^-_{i+\frac{1}{2}} = \frac{1}{2} \left(\tau^x_{i+\frac{1}{2}} - i \tau^y_{i+\frac{1}{2}}\right).
\end{align}
It is straightforward to verify that this qubit representation exactly preserves the gauge algebra within the truncated subspace $[\frac{1}{2}(1-\tau^z), \tau^-] = \tau^-$. Under this mapping, the electric field energy becomes $H_E \propto \sum_i (1 - 2\tau^z + \tau^z\tau^z)$. Dropping the constant energy shifts, we obtain $H_E = \sum_i \tau^z_{i+\frac{1}{2}}$.

\subsection{Jordan-Wigner Transformation}
To map the fermionic matter fields to qubit degrees of freedom (denoted by Pauli matrices $S$), we apply the Jordan-Wigner (JW) transformation. The fermionic operators are expressed as
\begin{align}
\psi_i &= \frac{1}{2} \left(\prod_{j<i} S^z_j\right) (S^x_i - i S^y_i) \equiv \left(\prod_{j<i} S^z_j\right) \sigma^-_i.
\end{align}
Using this transformation, the mass term trivially maps to the $Z$-basis
\begin{align}
\psi^\dagger_i \psi_i = \sigma^+_i \sigma^-_i = \frac{1}{2} \left(1 + S^z_i\right).
\end{align}
The kinetic hopping term incorporates both the JW strings and the gauge link operators. Because $\sigma^-_i$ and $S^z_i$ anticommute, the JW string between adjacent sites $i$ and $i+1$ simplifies cleanly, transforming the fermionic hopping $H_F$ into the Pauli interaction
\begin{equation}
\begin{aligned}
H_F = \sum_i \Big[ &\tau_{i+\frac{1}{2}}^x \left(S^x_{i+1}S^x_i + S^y_{i+1} S^y_{i} \right)+ \tau_{i+\frac{1}{2}}^y\left(S^y_{i+1} S^x_i - S^x_{i+1}S^y_{i} \right) \Big].
\end{aligned}
\end{equation}
Under periodic boundary conditions, the kinetic term connecting the first and last site ($1$ and $L$) must include the full JW string traversing the entire chain, yielding the boundary Hamiltonian $H_{F,\partial}$ with the string operator $S_{1\to L} = \prod_{i=1}^{L-1} S^z_i$ as presented in the main text. Finally, substituting the mapped operators into the Gauss's law yields the qubit-compatible generator
\begin{align}
G_i = \frac{1}{2} \left(\tau^z_{i-1/2} - \tau^z_{i+1/2} + S^z_i  +(-1)^i\right),
\end{align}
completing the fully compiled spin Hamiltonian used in our Trotterized circuits.

\section{Simplification of the Error channel via Symmetries}
\label{error_simplification}
Consider a target ideal unitary evolution $U = e^{-i H t}$. We can introduce a set of local covariant frame transformations $\mathcal{S} = \{S_k\}$, where $S_k$ are chosen based on the underlying symmetries of the target Hamiltonian (for simplicity, assumed here to be drawn from the Pauli group, $S_k = S_k^\dagger$, $S_k^2 = I$). For a given simulation, we transform the initial state, the circuit, and the observable as follows
\begin{equation}
    \rho \to S_k \rho S_k, \quad  \mathcal{O} \to S_k \mathcal{O} S_k.
\end{equation}
where $\rho$ is the initial state density matrix. 
At the ideal algorithmic level, the expectation value remains strictly invariant for any symmetry operation $S_k \in \mathcal{S}$
\begin{equation}
    \langle \mathcal{O} \rangle_k = \text{Tr}\left[ (S_k \mathcal{O} S_k)  U (S_k \rho S_k)  U^\dagger \right] = \text{Tr}[\mathcal{O} U \rho U^\dagger].
\end{equation}
Thus, averaging over the symmetry set $\mathcal{S}$ does not change the ideal outcome of the quantum simulation. 

However, at the physical hardware level, the operations are governed by an open quantum system dynamics described by the Lindblad master equation. The noisy evolution is generated by $\mathcal{L}(\rho) = -i[H, \rho] + \mathcal{D}(\rho)$, where the error channel $\mathcal{D}(\rho)$ for general Markovian errors is expanded in the Pauli basis $\{P_i\}$
\begin{equation}
    \mathcal{D}(\rho) = \sum_{i,j} \gamma_{ij} \left( P_i \rho P_j - \frac{1}{2}\{P_j P_i, \rho\} \right),
\end{equation}
where $\gamma_{ij}$ is the positive semi-definite decoherence matrix. In general, physical devices exhibit complex, non-diagonal cross-terms ($\gamma_{ij}$ for $i \neq j$), which severely complicates error mitigation techniques.

When we apply the symmetry frame transformation $S_k$, the physical error operator transforms rather than the state itself. The effective averaged dissipator over the chosen symmetry set $\mathcal{S}$ becomes
\begin{equation}
    \bar{\mathcal{D}}(\rho) = \frac{1}{|\mathcal{S}|} \sum_{S_k \in \mathcal{S}} S_k \mathcal{D}(S_k \rho S_k) S_k.
\end{equation}
Since $S_k$ either commutes or anticommutes with the Pauli operators $P_i$, we have $S_k P_i S_k = (-1)^{f(S_k, P_i)} P_i$, where $f(S_k, P_i) \in \{0, 1\}$. The effective decoherence matrix elements become
\begin{equation}
    \bar{\gamma}_{ij} = \gamma_{ij} \left( \frac{1}{|\mathcal{S}|} \sum_{S_k \in \mathcal{S}} (-1)^{f(S_k, P_i) + f(S_k, P_j)} \right).
\end{equation}
If $P_i \neq P_j$, evaluating the simulation across appropriately chosen symmetric frames ensures that the summation over the phase factors exactly vanishes. Consequently, the off-diagonal cross-terms are completely suppressed ($\bar{\gamma}_{ij} = 0$ for $i \neq j$), reducing the error channel to a much simpler Pauli or depolarizing channel.

\subsection{Examples: Single-Qubit and Two-Qubit Errors}
Unlike standard randomized compiling, which requires inserting deep sequences of random gates, our approach achieves error simplification natively by exploiting the inherent symmetries of the target system. For instance, consider the $XXZ$ spin model, which is invariant under global $Z$-axis rotations. Simulating the dynamics in a $Z$-rotated frame yields identical ideal expectation values, but crucially, it transforms the physical error channels.

\paragraph{Single-Qubit Errors:}

Consider a generic single-qubit noise model where the error channel is
\begin{align}
\mathcal{E}(\rho)
= \gamma_{X} X \rho X + \gamma_{Y} Y \rho Y + \gamma_{Z} Z \rho Z.
\end{align}
Instead of generic Pauli twirling, we can choose the averaging set based on the $XXZ$ model's global $Z$-rotation symmetry $\mathcal{S} = \{Z^{\otimes N}\}$, or the Heisenberg model's global SU(2) symmetry.

For any local site, conjugation by $e^{-i \theta Z}$ yields 
\begin{align}
e^{+i \theta Z} \begin{pmatrix}
X\\
Y
\end{pmatrix} e^{-i \theta Z}
=\begin{pmatrix}
\cos 2\theta &\sin 2\theta \\
-\sin 2\theta &\cos 2\theta \\
\end{pmatrix}
\begin{pmatrix}
X\\
Y
\end{pmatrix}. 
\end{align}
When we average the measurement outcomes from the original and the $Z$-rotated symmetry frames, the effective error channel simplifies to
\begin{align}
\gamma_{X} X \rho X + \gamma_{Y} Y \rho Y + \gamma_{Z} Z \rho Z
~\to~
\gamma_{X} (X \rho X + Y \rho Y) + \gamma_{Z} Z \rho Z,
\end{align}
and averaging over arbitrary global $SU(2)$ rotations simplifies more general error-channel tensors even further.

\paragraph{Two-Qubit Errors:}
{Now consider a generic two-qubit noise model where the error channel is given by
\begin{align}
\gamma_{ij} S^i S^j \rho S^j S^i ,
\end{align}
Suppose the Hamiltonian exhibits global $X$-, $Y$-, $Z$- axis rotation symmetries like Heisenberg model, 
By taking covariant measurement outcomes in rotated frames, specifically, frames rotated by $e^{-i X \theta}$, $e^{-i Y \theta}$, $e^{-i Z \theta}$, we can simplify the error tensor:
\begin{align}
\begin{pmatrix}
\gamma_{XX}&\gamma_{XY}&\gamma_{XZ}\\
\gamma_{YX}&\gamma_{YY}&\gamma_{YZ}\\
\gamma_{ZX}&\gamma_{ZY}&\gamma_{ZZ}
\end{pmatrix}
\to
\begin{pmatrix}
\gamma_{d}&\gamma_{1}&\gamma_{2}\\
\gamma_{2}&\gamma_{d}&\gamma_{1}\\
\gamma_{1}&\gamma_{2}&\gamma_{d}
\end{pmatrix},
\end{align}
where the simplified components are defined as:
\begin{align}
\gamma_{d} &= \frac{1}{3}\left(\gamma_{XX}+ \gamma_{YY}+ \gamma_{ZZ}\right),\\
\gamma_{1} &=\frac{1}{3}\left( \gamma_{XY}+ \gamma_{YZ}+ \gamma_{ZX}\right),\\
\gamma_{2} &=\frac{1}{3}\left( \gamma_{YX}+ \gamma_{XZ}+ \gamma_{ZY}\right).
\end{align}
This simplified error channel significantly lowers the sampling overhead required for advanced quantum error mitigation protocols, such as zero-noise extrapolation or probabilistic error cancellation.

Furthermore, this concept can be generalized to broader classes of Hamiltonians. Consider starting from an initial state and measuring $\sum_{i} Z_i$ under the anisotropic Heisenberg $H_{XXZ}$ Hamiltonian, which possesses a global $Z$-axis rotation symmetry. The physical outcome of this measurement is equivalent to measuring $\sum_i X_i $ under a rotated Hamiltonian $H_{XZZ}$ with a global $X$-axis rotation symmetry, provided the initial state is appropriately rotated. While the noiseless measurement outcomes are identical, the underlying physical noise channels are generally not invariant under this transformation. Therefore, averaging these symmetry-related measurement outcomes allows us to further simplify the effective error channels by symmetrizing both the algorithmic and physical noise contributions.}

\end{document}